\documentclass[conference]{IEEEtran}
\usepackage{cite}
\usepackage{amsmath}
\usepackage{amssymb}
\usepackage{amsthm}
\usepackage{arydshln}
\usepackage{mathtools}
\usepackage{mathrsfs}
\usepackage{algorithmic}
\usepackage{graphicx}
\usepackage{epsfig,graphics}
\usepackage{epstopdf}
\usepackage{subfigure}
\usepackage{enumitem}

\usepackage{multirow,ragged2e}
\usepackage{bm}
\usepackage{fixltx2e}
\usepackage[utf8]{inputenc}
\usepackage[english]{babel}
\usepackage{filecontents}
\theoremstyle{theorem}

\theoremstyle{lemma}

\newtheorem{rem}{Remark}

\newcommand{\fig}[1]{Fig.~\ref{#1}}
\newcommand{\figs}[2]{Fig.~\ref{#1}-\ref{#2}}

\newcommand{\sect}[1]{Section~\ref{#1}}
\newcommand{\eq}[1]{Equation~(\ref{#1})}
\newcommand{\eqs}[2]{Equations~(\ref{#1})-(\ref{#2})}
\newcommand{\eqsa}[2]{Equations~(\ref{#1}) and~(\ref{#2})}
\newcommand{\eqsthree}[3]{Equations~(\ref{#1}),~(\ref{#2}) and~(\ref{#3})}

\newcommand{\fbm}[1]{\mathbf{#1}}
\newcommand{\tbm}[1]{\fbm{#1}^\mathsf{T}}
\newcommand{\tfbm}[1]{\bm{#1}^\mathsf{T}}
\newcommand{\ibm}[1]{\fbm{#1}^{-1}}

\newcommand{\hbm}[1]{\hat{\fbm{#1}}}
\newcommand{\thbm}[1]{\hat{\fbm{#1}}^\mathsf{T}}
\newcommand{\tilbm}[1]{\tilde{\fbm{#1}}}

\newcommand{\dottbm}[1]{\dot{\fbm{#1}}^\mathsf{T}}
\newcommand{\dotbm}[1]{\dot{\fbm{#1}}}
\newcommand{\dotfbm}[1]{\dot{\bm{#1}}}
\newcommand{\hatdotbm}[1]{\hat{\dot{\fbm{#1}}}}
\newcommand{\dothatbm}[1]{\dot{\hat{\fbm{#1}}}}
\newcommand{\dottilbm}[1]{\dot{\tilde{\fbm{#1}}}}

\newcommand{\tildottbm}[1]{\tilde{\dotbm{#1}}^\mathsf{T}}
\newcommand{\tildotbm}[1]{\tilde{\dotbm{#1}}}

\newcommand{\ddotbm}[1]{\ddot{\fbm{#1}}}

\begin{document}

\title{Globally Stable Output Feedback Synchronization of Teleoperation with Time-Varying Delays}

\author{\IEEEauthorblockN{Yuan Yang}
\IEEEauthorblockA{Department of Mechanical Engineering\\
University of Victoria\\
Victoria, British Columbia 250-853-3220\\
Email: yangyuan@uvic.ca}
\and
\IEEEauthorblockN{Daniela Constantinescu}
\IEEEauthorblockA{Department of Mechanical Engineering\\
University of Victoria\\
Victoria, British Columbia 250-721-6040\\
Email: danielac@uvic.ca}
\and
\IEEEauthorblockN{Yang Shi}
\IEEEauthorblockA{Department of Mechanical Engineering\\
University of Victoria\\
Victoria, British Columbia 250-853-3178\\
Email: yshi@uvic.ca}
}

\maketitle

\begin{abstract}
This paper presents a globally stable teleoperation control strategy for systems with time-varying delays that eliminates the need for velocity measurements through novel augmented Immersion and Invariance velocity observers. The new observers simplify a recent constructive Immersion and Invariance velocity observer to achieve globally convergent velocity estimation with only $n+2$ states, where $n$ is the number of degrees of freedom of the master and slave robots. They introduce dynamic scaling factors to accelerate the speed of convergence of the velocity estimates and, thus, to limit the energy generated by the velocity estimation errors and to guarantee sufficient estimate-based damping injection to dissipate the energy generated by the time-varying delays. The paper shows that Proportional plus damping control with the simplified and augmented Immersion and Invariance-based velocity observers can synchronize the free master and slave motions in the presence of time-varying delays without using velocity measurements. Numerical results illustrate the estimation performance of the new observers and the stability of a simulated two degrees-of-freedom nonlinear teleoperation system with time-varying delays under the proposed output feedback Proportional plus damping control.
\end{abstract}

\IEEEpeerreviewmaketitle

\section{Introduction}\label{sec:introduction}
In the quest for guaranteed stability in the presence of communication delays, bilateral teleoperation research has developed several passivity-based control strategies~\cite{Nuno2011}. Through Lyapunov-Krasovskii analysis, Proportional-Derivative plus damping~(PD+d) control has been shown to stabilize bilateral teleoperation in the sense of bounded velocities of, and bounded position error between, the master and slave robots if the asymmetric communication delays are constant and the user and environment are passive~\cite{Lee2006TRO}. A simpler Proportional plus damping~(P+d) strategy has also been proven to stabilize bilateral teleoperation whether the delays are constant~\cite{Nuno2008TRO} or time-varying~\cite{Nuno2009IJRR, Hua2010TRO}.

Damping injection requires velocity measurements, but most commercial robots are not equipped with velocity sensors. Velocity estimation through carefully designed velocity observers~\cite{Sarras2016IJRNC} has been used to inject damping in teleoperation with only position measurements. Because nonlinear bilateral teleoperators are semi-autonomous systems, one challenge facing the observer design is the need to guarantee the convergence of velocity estimates without assuming bounded master and slave velocities. Recently, the Immersion and Invariance~(I\&I) velocity observer~\cite{Astolfi2009CDC,Astolfi2010Automatica} has been proven globally exponentially convergent and has been used for trajectory tracking in Euler-Lagrange systems~\cite{Romero2013ACC,Romero2015TAC}. A constructive version of it~\cite{Stamnes2011Automatica}, with simpler dynamics and computed based on the exact solution of a partial differential equation~(PDE), has also been employed for output feedback tracking control of Euler-Lagrange dynamics\cite{Stamnes2011IFAC}. A second challenge when using velocity observers in teleoperation systems is that the estimation errors inject deleterious energy in the closed-loop system. Practically, damping injection based on velocity estimates dissipates the energy induced by the delays but generates energy through estimation errors. To not threaten stability, the observers must converge sufficiently fast to create less energy through estimation errors than that they can dissipate.

This paper investigates the globally stable output feedback synchronization of bilateral teleoperation systems with time-varying delays. Like in~\cite{Nuno2009IJRR}, the proposed observer-based P+d control guarantees stable teleoperation in the sense of bounded velocities of, and position error between, the master and slave robots. In the absence of operator and environment forces, the velocities of, and position error between, the two robots asymptotically converge to zero. The main contributions of this work are:
\begin{enumerate}
\item
Compared to the I\&I observers with $3n+1$ states in~\cite{Astolfi2009CDC,Astolfi2010Automatica} and with $2n+2$ states in~\cite{Stamnes2011Automatica}, the I\&I observers in this paper guarantee exponential convergence of the velocity estimates with only $n+2$ states, where $n$ is the number of degrees of freedom of the robots. 
\item
Compared to~\cite{Astolfi2009CDC,Astolfi2010Automatica,Stamnes2011Automatica,Cao2017TAC}, the I\&I observers in this paper do not require the state transformations. Their dynamics involve robot-independent scalar gains instead of robot-dependent, analytically derived matrix gains and their derivatives~\cite{Astolfi2009CDC,Astolfi2010Automatica,Stamnes2011Automatica}, leading to a much simpler observer design procedure.
\item
In~\cite{Cao2017TAC}, the designed observer is applicable for only $2$-link revolute robot manipulators. However, the observer in this paper is designed for any $n$-DOF nonredundant robot arms, and is thus more general than~\cite{Cao2017TAC}. 
\item
To kepp the system stable in the presence of time-varying delays, in the observer dynamics, the estimation gains $k_{xi}$ in $\dotfbm{\xi}_{i}$ are updated according the dynamic scaling factors $r_{i}$. This augmentation of the observer dynamics increases the speed of convergence of velocity estimation and thus, limits the energy injected by the estimation errors within a range that the observers themselves can dissipate. The integration of the new observers into conventional P+d control~\cite{Nuno2009IJRR} leads to rigorously provable global stability of output feedback synchronization of teleoperation systems with time-varying delays.
\end{enumerate} 

\section{System dynamics}\label{sec:system dynamics}

The nonlinear dynamics of a teleoperator with $n$-degrees-of-freedom~($n$-DOF) master and slave robots with serial links and revolute joints, are:
\begin{equation}\label{equ1}
\fbm{M}_{i}(\fbm{q}_{i})\ddot{\fbm{q}}_{i}+\fbm{C}_{i}(\fbm{q}_{i},\dot{\fbm{q}}_{i})\dot{\fbm{q}}_{i}+\fbm{g}_{i}(\fbm{q}_{i})=\bm{\tau}_{he}+\bm{\tau}_{i}\textrm{,}
\end{equation}
where: the subscripts $i=m,s$ indicate the master and slave robots, respectively; $\fbm{q}_{i}$, $\dot{\fbm{q}}_{i}$ and $\ddot{\fbm{q}}_{i}$ are the joint position, velocity and acceleration vectors; $\fbm{M}_{i}(\fbm{q}_{i})$ and $\fbm{C}_{i}(\fbm{q}_{i},\dot{\fbm{q}}_{i})$ are the joint space matrices of inertia and of Coriolis and centrifugal effects; $\fbm{g}_{i}(\fbm{q}_{i})$ are the gravity joint torques; $\bm{\tau}_{he}=\bm{\tau}_{h}$ if $i=m$ and $\bm{\tau}_{he}=\bm{\tau}_{e}$ if $i=s$ are the joint torques due to the hand and environment forces, respectively; $\bm{\tau}_{i}$ are the control joint torques. For shorter notation and ease of reading, the dependence on states of the inertia matrices, of the Coriolis and cetrifugal effects matrices and of the gravity vectors is dropped from notation hereafter, i.e., $\fbm{M}_{i}$, $\fbm{C}_{i}$ and $\fbm{g}_{i}$ are used in place of $\fbm{M}_{i}(\fbm{q}_{i})$, $\fbm{C}_{i}(\fbm{q}_{i},\dot{\fbm{q}}_{i})$ and $\fbm{g}_{i}(\fbm{q}_{i})$. 

The stability analysis in \sect{sec:stability analysis} relies on the following:
\begin{itemize}
\item properties of the nonlinear dynamics in \eq{equ1}: 

\begin{enumerate}
\renewcommand{\theenumi}{P\arabic{enumi}}
\item\label{itm:P1} 
The inertia matrices $\fbm{M}_{i}$ are uniformly lower and upper bounded, i.e., $\exists \lambda_{i1}>0, \lambda_{i2}>0$ for $i=m,s$ such that: $\fbm{0}\prec\lambda_{i1}\fbm{I}\preceq \fbm{M}_{i}\preceq \lambda_{i2}\fbm{I}\prec\infty$.\par

\item\label{itm:P2} 
$\dot{\fbm{M}}_{i}-2\fbm{C}_{i}$ are skew symmetric.\par

\item\label{itm:P3} 
There exist $c_{i}>0$ such that $\|\fbm{C}_{i}(\fbm{q}_{i},\fbm{x})\fbm{y}\|\leq c_{i}\|\fbm{x}\|\|\fbm{y}\|,\quad \forall \fbm{q}_{i}, \fbm{x}, \fbm{y}$.\par
\end{enumerate}

\item assumptions on the communication delays and on the hand and environment torques:
\begin{enumerate}
\renewcommand{\theenumi}{A\arabic{enumi}}
\item\label{itm:A1} 
The forward $d_{m}$ and backward $d_{s}$ communication delays are positive with known finite bounds: $0\leq d_{i}\leq \overline{d}_{i}$, $i=m,s$.

\item\label{itm:A2} 
The human operator and environment are passive: $E_{h}-\int^{t}_{0}\dottbm{q}_{m}\bm{\tau}_{h}d\xi\geq 0$, and $E_{e}-\int^{t}_{0}\dottbm{q}_{s}\bm{\tau}_{e}d\xi\geq 0$, where $E_{h}$ and $E_{e}$ are positive constants.
\end{enumerate}

\end{itemize}

\section{Observer-Based Controller Design}\label{sec:observer-based controller design}  

The master and slave dynamics in~\eq{equ1} can be written:
\begin{equation}\label{equ2}
\begin{aligned}
\dotbm{x}_{i}=&\ibm{M}_{i}(\fbm{y}_{i})\big[-\fbm{C}_{i}(\fbm{y}_{i},\fbm{x}_{i})\fbm{x}_{i}-\fbm{g}_{i}(\fbm{y}_{i})+\fbm{u}_{i}\big]\textrm{,}
\end{aligned}
\end{equation}
where: $\fbm{y}_{i}=\fbm{q}_{i}$, $\fbm{x}_{i}=\dotbm{q}_{i}$, $\fbm{u}_{m}=\bm{\tau}_{m}+\bm{\tau}_{h}$ and $ \fbm{u}_{s}=\bm{\tau}_{s}+\bm{\tau}_{e}$ with $i=m,s$. Then, the new augmented I$\&$I observers on the master and slave sides are designed as follows:
\begin{equation}\label{equ3}
\begin{aligned}
\hbm{x}_{i}=&\bm{\xi}_{i}+k_{xi}(r_{i},\hat{\sigma}_{i})\fbm{y}_{i}\textrm{,}\\
\dot{\bm{\xi}}_{i}=&\fbm{f}_{i}-k_{xi}(r_{i},\hat{\sigma}_{i})\hbm{x}_{i}-\dot{k}_{xi}(r_{i},\dot{r}_{i},\dot{\hat{\sigma}}_{i})\fbm{y}_{i}\textrm{,}\\
\dot{r}_{i}=&-\frac{k_{r}}{2}(r_{i}-c_{ri})+\frac{k_{r}}{4\lambda_{i1}}c^{2}_{i}|\tilde{\sigma}|r_{i}\textrm{,}\\
\dot{\hat{\sigma}}_{i}=&\text{Proj}_{\hat{\sigma}_{i}}\Big(2\left[ \thbm{x}_{i}\fbm{f}_{i}+k_{\sigma i}(\hbm{x}_{i},r_{i},\hat{\sigma}_{i})\tilde{\sigma}_{i}\right]\Big)\textrm{,}
\end{aligned}
\end{equation}
where $i=m,s$ and:
\begin{align*}
&\fbm{f}_{i}=\ibm{M}_{i}(\fbm{y}_{i})\big[-\fbm{C}_{i}(\fbm{y}_{i},\hbm{x}_{i})\hbm{x}_{i}-\fbm{g}_{i}(\fbm{y}_{i})+\fbm{u}_{i}\big]\textrm{,}\\
&\textrm{Proj}_{\hat{\sigma}_{i}}(\tau)=\begin{cases}\tau\textrm{,} \quad& \hat{\sigma}_{i}>0\ \text{or}\ \tau\geq 0\\
(1-c_{\sigma i}(\hat{\sigma}_{i}))\tau\textrm{,} \quad& -\epsilon_{i}\leq\hat{\sigma}_{i}\leq 0, \tau<0\end{cases}\textrm{,}\\
&k_{xi}(r_{i},\hat{\sigma}_{i})=\frac{1}{\lambda_{i1}}\left[\frac{2}{k_{r}}+\frac{k_{r}}{4}\left(3\lambda_{i2}+c^{2}_{i}\hat{\sigma}_{i}\right)+\frac{1}{4}\alpha_{i}k_{i}r^{2}_{i}\right]\textrm{,}\\
&k_{\sigma i}(\hbm{x}_{i},r_{i},\hat{\sigma}_{i})=\frac{k_{r}}{16}\left(\frac{c^{4}_{i}r^{2}_{i}}{\lambda^{2}_{i1}}+4k^{2}_{xi}(r_{i},\hat{\sigma}_{i})\|\hbm{x}_{i}\|^{2}r^{2}_{i}+2\right)\textrm{,}\\
&\dot{k}_{xi}(r_{i},\dot{r}_{i},\dot{\hat{\sigma}}_{i})= \frac{1}{4\lambda_{i1}}\left(2\alpha_{i}k_{i}r_{i}\dot{r}_{i}+k_{r}c^{2}_{i}\dot{\hat{\sigma}}_{i}\right)\textrm{,}
\end{align*}
with $c_{ri}$, $k_{r}$ and $k_{i}$ being positive constants and $c_{\sigma i}(\hat{\sigma}_{i})=\text{min}\{1,\frac{-\hat{\sigma}_{i}}{\epsilon_{i}}\}$ for $0<\epsilon_{i}<1$. The observer output $\hbm{x}_{i}$ and the observer state $\hat{\sigma}_{i}$ are estimates of $\fbm{x}_{i}$ and $\sigma_{i}=\|\hbm{x}_{i}\|^{2}$, respectively, with corresponding estimation errors $\tilbm{x}_{i}=\fbm{x}_{i}-\hbm{x}_{i}$ and $\tilde{\sigma}_{i}=\sigma_{i}-\hat{\sigma}_{i}$. By design, $[c_{ri},+\infty)$ is an invariant set of $r_{i}$, i.e., $r_{i}(t)\geq c_{ri}>0$ $\forall t\geq 0$ if $r_{i}(0)\geq c_{ri}>0$.       

\begin{rem}
\normalfont By sidesteping the need to estimate $\fbm{y}$, the observer in this paper offers several advantages compared to~\cite{Stamnes2011Automatica}: reduced dimension $n+2$, instead of $2n+2$; a scalar gain $k_{xi}(r_{i},\hat{\sigma})$ dependent on the dynamic scaling $r_i$, instead of a matrix gain $\fbm{K}_{x}(\hat{\sigma},\hbm{y})$ dependent on $\fbm{y}$, and thus, no need for analytical solutions of $-\frac{\partial\fbm{K}_{x}}{\partial\hbm{y}}\dothatbm{y}-\frac{\partial\fbm{K}_{x}}{\partial\hat{\sigma}}\dot{\hat{\sigma}}$ in $\dotfbm{\xi}$; and simpler dynamics of the scaling $r_i$, dependent on $\frac{k_{r}}{4\lambda_{i1}}c^{2}_{i}|\tilde{\sigma}|$ instead of the non-smooth functions $\bar{\Delta}_{y}(\fbm{y},\hbm{y},\hat{\sigma})\|\tilbm{y}\|+\bar{\Delta}_{\sigma}(\fbm{y},\hbm{x},\hat{\sigma})\|\tilde{\sigma}\|$.
\end{rem}

\begin{rem}
\normalfont To convert the system dynamics in~\eq{equ1} to a form suitable for use in the observer design, the I\&I observers in~\cite{Astolfi2009CDC,Astolfi2010Automatica,Stamnes2011Automatica,Cao2017TAC} require several matrix decoupling, inversion, differentiation and multiplication operations to be performed analytically. Hence, their practical implementation is not trivial. In contrast, the I\&I observers in this paper use the converted system dynamics in~\eq{equ2} and need only the inverse inertia matrices $\ibm{M}_{i}$, which can be computed online. This contributes to a simpler design procedure.
\end{rem}

Given the velocity observers in \eq{equ3}, conventional P+d teleoperation control can be implemented using the velocity estimates $\hatdotbm{q}_{i}=\hbm{x}_{i}$ instead of the velocity measurements:
\begin{equation}\label{equ4}
\bm{\tau}_{i}=-p(\fbm{q}_{i}-\fbm{q}_{jd})-k_{i}\hatdotbm{q}_{i}+\fbm{g}_{i}\textrm{,}
\end{equation}
where: $i,j=m,s$ and $j\neq i$; $p$ and $k_{i}$ are positive constant gains; $\fbm{q}_{jd}=\fbm{q}_{j}(t-d_{j}(t))$ are delayed position signals; $\fbm{g}_{i}$ are gravitational torque compensation terms. The control terms proportional to the position errors $-p(\fbm{q}_{i}-\fbm{q}_{jd})$ synchronize the master and slave robots. The damping injection terms $-k_{i}\hatdotbm{q}_{i}$ dissipate the energy induced by the time-varying delays. Because damping injection is based on estimated velocities, the estimation errors can also inject energy in the closed-loop system and potentially lead to instability. The design methodology in this paper is then to use the new I\&I observers to consume the energy generated by the estimation errors. To this end, the master and slave velocity observers have more general dynamics than in~\cite{Stamnes2011Automatica}, with gains $k_{xi}(r_{i},\hat{\sigma}_{i})$ that depend not only on $\hat{\sigma}_{i}$ but also on the scaling factors $r_{i}$. The dependence of the dynamics of the velocity estimates on the dynamic scaling factors is critical for the rigorous proof of the global stability of the system. Essentially, the proof suggests that adjusting the speed of convergence of the velocity estimates based on the robot and observer states determines the dissipativity of the designed observers.

\section{Stability Analysis}\label{sec:stability analysis}  

The stability of a bilateral teleoperation system in closed-loop with a P+d controller that uses position measurements and the velocity estimates provided by the augmented constructive I\&I observers in \eq{equ3} is analyzed using the following Lyapunov-like functional:
\begin{equation}\label{equ5}
V=\sum^{3}_{k=1}V_{k}+V_{o}
\textrm{,}
\end{equation}
where:
\begin{align*}
V_{1}=&\frac{1}{2}\sum_{i=m,s}\dottbm{q}_{i}\fbm{M}_{i}\dotbm{q}_{i}+\frac{p}{2}(\fbm{q}_{m}-\fbm{q}_{s})^\mathsf{T}(\fbm{q}_{m}-\fbm{q}_{s})\textrm{,}\\
V_{2}=&-\int^{t}_{0}\dottbm{q}_{m}\bm{\tau}_{h}d\xi-\int^{t}_{0}\dottbm{q}_{s}\bm{\tau}_{e}d\xi+E_{h}+E_{e}\textrm{,}\\
V_{3}=&\sum_{i=m,s}\omega_{i}\int^{0}_{-\overline{d}_{i}}\int^{t}_{t+\theta}\dottbm{q}_{i}\dotbm{q}_{i}d\xi d\theta\textrm{,}\\
V_{o}=&\frac{1}{4}\sum_{i=m,s}\Big\{2\tbm{\bm{\eta}}_{i}\fbm{M}_{i}\bm{\eta}_{i}+2(r_{i}-c_{ri})^{2}+\tilde{\sigma}^{2}_{i}\Big\}\textrm{.}
\end{align*}
In \eq{equ5}: $V_{1}$ is the sum of the kinetic energies of the master and slave robots and the potential energy stored in the Proportional control; $V_{2}$ is the energy input by the user and environment, and is positive based on the assumption \ref{itm:A2} that both are passive; $V_{3}$ is a measure of the energy generated by the time-varying delays, and is non-negative for positive $\omega_{i}$ ; and $V_{o}$ is used to prove global convergence of the velocity estimates and that the observers themselves dissipate the harmful energy generated by the estimation errors, with $\bm{\eta}_{i}=\frac{\tilbm{x}_{i}}{r_{i}}$ being scaled versions of the velocity estimation errors. 

From \eqsa{equ1}{equ4} and property \ref{itm:P2}, the summation of the derivatives of $V_{1}$ and $V_{2}$ is:
\begin{align*}
\sum^{2}_{k=1}\dot{V}_{k}
=&\frac{1}{2}\sum_{i=m,s}\dottbm{q}_{i}\dotbm{M}_{i}\dotbm{q}_{i}+\sum_{i=m,s}\dottbm{q}_{i}\fbm{M}_{i}\ddotbm{q}_{i}\\
&+p(\fbm{q}_{m}-\fbm{q}_{s})^\mathsf{T}(\dotbm{q}_{m}-\dotbm{q}_{s})\\
=&\sum_{i=m,s}\left\{p\dottbm{q}_{i}(\fbm{q}_{i}-\fbm{q}_{j})-p\dottbm{q}_{i}(\fbm{q}_{i}-\fbm{q}_{jd})-k_{i}\dottbm{q}_{i}\hat{\dotbm{q}}_{i}\right\}
\end{align*}
\begin{equation}\label{equ6}
\begin{aligned}
=&\sum_{i,j=m,s}\left\{k_{i}\dottbm{q}_{i}\tildotbm{q}_{i}-k_{i}\dottbm{q}_{i}\dotbm{q}_{i}-p\dottbm{q}_{i}(\fbm{q}_{j}-\fbm{q}_{jd})\right\}\\
\leq&\sum_{i,j=m,s}\Big\{\frac{\alpha_{i}k_{i}}{4}\tildotbm{q}_{i}\tildotbm{q}_{i}-p\dottbm{q}_{i}\int^{t}_{t-d_{j}}\dotbm{q}_{j}d\xi-(1-\frac{1}{\alpha_{i}})k_{i}\dottbm{q}_{i}\dotbm{q}_{i}\Big\}\textrm{,}
\end{aligned}
\end{equation}
where $k_{i}\dottbm{q}_{i}\tildotbm{q}_{i}\leq\frac{k_{i}}{\alpha_{i}}\dottbm{q}_{i}\dotbm{q}_{i}+\frac{\alpha_{i}k_{i}}{4}\tildotbm{q}_{i}\tildotbm{q}_{i}$ with $\tildotbm{q}_{i}=\dotbm{q}_{i}-\hatdotbm{q}_{i}$ and $\alpha_{i}>1$ have been used.

After bounding the derivative of $V_{3}$ by:
\begin{equation}\label{equ7}
\begin{aligned}
\dot{V}_{3}= &\sum_{i=m,s}\overline{d}_{i}\omega_{i}\dottbm{q}_{i}\dotbm{q}_{i}-\omega_{i}\int^{t}_{t-\overline{d}_{i}}\dottbm{q}_{i}\dotbm{q}_{i}d\xi\\
\leq &\sum_{i=m,s}\overline{d}_{i}\omega_{i}\dottbm{q}_{i}\dotbm{q}_{i}-\omega_{i}\int^{t}_{t-d_{i}}\dottbm{q}_{i}\dotbm{q}_{i}d\xi\textrm{,}
\end{aligned}
\end{equation}
algebraic manipulations using Lemma 1 in~\cite{Hua2010TRO} lead to:
\begin{equation}\label{equ8}
-p\dottbm{q}_{j}\int^{t}_{t-d_{i}}\dotbm{q}_{i}d\xi-\omega_{i}\int^{t}_{t-d_{i}}\dottbm{q}_{i}\dotbm{q}_{i}d\xi\leq\frac{\overline{d}_{i}p^{2}}{4\omega_{i}}\dottbm{q}_{j}\dotbm{q}_{j}\textrm{,}
\end{equation}
where $i,j=m,s$ and $i\neq j$. After substitution from \eq{equ8}, the sum of \eqs{equ6}{equ7} yields:
\begin{equation}\label{equ9}
\sum^{3}_{k=1}\dot{V}_{k}\leq\sum_{i=m,s}\rho_{i}\dottbm{q}_{i}\dotbm{q}_{i}+\frac{\alpha_{i}k_{i}}{4}\tildottbm{q}_{i}\tildotbm{q}_{i}
\end{equation}
with $\rho_{i}=\overline{d}_{i}\omega_{i}+\frac{\overline{d}_{j}p^{2}}{4\omega_{j}}-(1-\frac{1}{\alpha_{i}})k_{i}$.

\begin{rem}
\normalfont In \eq{equ6}, $\frac{\alpha_{i}k_{i}}{4}\tildotbm{q}_{i}\tildotbm{q}_{i}-p\dottbm{q}_{i}\int^{t}_{t-d_{j}}\dotbm{q}_{j}d\xi$ show that the velocity estimation errors and the time-varying delays are the two sources of possible energy injection and, hence, of instability in the closed-loop teleoperation system. \eq{equ8} indicates that the damping injected by the P+d controllers can dissipate the delay-induced energy $-p\dottbm{q}_{i}\int^{t}_{t-d_{j}}\dotbm{q}_{j}d\xi$. In contrast, \eq{equ9} implies that the P+d controllers cannot consume the energy created by the velocity estimation errors, $\frac{\alpha_{i}k_{i}}{4}\tildotbm{q}_{i}\tildotbm{q}_{i}$. The observers themselves need to dissipate this energy.
\end{rem}

From \eq{equ3}, the derivatives of $\hbm{x}_{i}$, $i=m,s$, are:
\begin{align*}
\dothatbm{x}_{i}=&\dot{\bm{\xi}}_{i}+\dot{k}_{xi}(r_{i},\dot{r}_{i},\dot{\hat{\sigma}}_{i})\fbm{y}_{i}+k_{xi}(r_{i},\hat{\sigma}_{i})\dotbm{y}_{i}\\
=&\fbm{f}_{i}-k_{xi}(r_{i},\hat{\sigma}_{i})\hbm{x}_{i}-\dot{k}_{xi}(r_{i},\dot{r}_{i},\dot{\hat{\sigma}}_{i})\fbm{y}_{i}\\
&+\dot{k}_{xi}(r_{i},\dot{r}_{i},\dot{\hat{\sigma}}_{i})\fbm{y}_{i}+k_{xi}(r_{i},\hat{\sigma}_{i})\dotbm{y}_{i}\\
=&\fbm{f}_{i}+k_{xi}(r_{i},\hat{\sigma}_{i})\tilbm{x}_{i}\textrm{,}
\end{align*}
because $\dotbm{y}_{i}=\fbm{x}_{i}$. Then, property~\ref{itm:P2} leads to the error dynamics:
\begin{align*}
\dottilbm{x}_{i}
=&\ibm{M}_{i}(\fbm{y}_{i})\big[-\fbm{C}_{i}(\fbm{y}_{i},\fbm{x}_{i})\fbm{x}_{i}+\fbm{C}_{i}(\fbm{y}_{i},\fbm{x}_{i})\hbm{x}_{i}\\
&-\fbm{C}_{i}(\fbm{y}_{i},\fbm{x}_{i})\hbm{x}_{i}+\fbm{C}_{i}(\fbm{y}_{i},\hbm{x}_{i})\hbm{x}_{i}\big]-k_{xi}(r_{i},\hat{\sigma}_{i})\tilbm{x}_{i}\\
=&-\ibm{M}_{i}(\fbm{y}_{i})\big[\fbm{C}_{i}(\fbm{y}_{i},\fbm{x}_{i})\tilbm{x}_{i}+\fbm{C}_{i}(\fbm{y}_{i},\tilbm{x}_{i})\hbm{x}_{i}\big]\\
&-k_{xi}(r_{i},\hat{\sigma}_{i})\tilbm{x}_{i}
\textrm{.}
\end{align*}
Given the derivatives of $\bm{\eta}_{i}$:
\begin{align*}
\dot{\bm{\eta}}_{i}=&\frac{1}{r_{i}}\dottilbm{x}_{i}-\frac{\dot{r}_{i}}{r^{2}_{i}}\tilbm{x}_{i}=-\ibm{M}_{i}(\fbm{y}_{i})\fbm{C}_{i}(\fbm{y}_{i},\fbm{x}_{i})\bm{\eta}_{i}-\frac{\dot{r}_{i}}{r_{i}}\bm{\eta}_{i}\\
&-k_{xi}(r_{i},\hat{\sigma}_{i})\bm{\eta}_{i}-\ibm{M}_{i}(\fbm{y}_{i})\fbm{C}_{i}(\fbm{y}_{i},\bm{\eta}_{i})\hbm{x}_{i}\textrm{,}
\end{align*}
the derivative of $V_{\eta}=\frac{1}{2}\sum_{i=m,s}\tfbm{\eta}_{i}\fbm{M}_{i}\bm{\eta}_{i}$ becomes
\begin{align*}
\dot{V}_{\eta}=&\frac{1}{2}\sum_{i=m,s}\left\{\tfbm{\eta}_{i}\dotbm{M}_{i}\bm{\eta}_{i}+2\tfbm{\eta}_{i}\fbm{M}_{i}\dotfbm{\eta}_{i}\right\}\\
=&-\sum_{i=m,s}\Big\{\tfbm{\eta}_{i}\fbm{C}_{i}(\fbm{y}_{i},\bm{\eta}_{i})\hbm{x}_{i}+k_{xi}(r_{i},\hat{\sigma}_{i})\tfbm{\eta}_{i}\fbm{M}_{i}(\fbm{y}_{i})\bm{\eta}_{i}\\
&\quad \quad \quad \quad +\frac{\dot{r}_{i}}{r_{i}}\tfbm{\eta}_{i}\fbm{M}_{i}(\fbm{y}_{i})\bm{\eta}_{i}\Big\}\textrm{,}
\end{align*}
where 
\begin{align*}
-\tfbm{\eta}_{i}\fbm{C}_{i}(\fbm{y}_{i},\bm{\eta}_{i})\hbm{x}_{i}\leq\frac{1}{k_{r}}\|\bm{\eta}_{i}\|^{2}+\frac{k_{r}c^{2}_{i}}{4}(\hat{\sigma}_{i}+|\tilde{\sigma}_{i}|)\|\bm{\eta}_{i}\|^{2}
\end{align*}
and
\begin{align*}
-\frac{\dot{r}_{i}}{r_{i}}\tfbm{\eta}_{i}\fbm{M}_{i}(\fbm{y}_{i})\bm{\eta}_{i}\leq\frac{k_{r}}{4}\left(2\lambda_{i2}-c^{2}_{i}|\tilde{\sigma}_{i}|\right)\|\bm{\eta}_{i}\|^{2}\textrm{.}
\end{align*}
Then, $\dot{V}_{\eta}$ can then be upper-bounded by
\begin{equation}\label{equ10}
\dot{V}_{\eta}\leq \sum_{i=m,s}\left[\frac{1}{k_{r}}+\frac{k_{r}}{4}(c^{2}_{i}\hat{\sigma}_{i}+2\lambda_{i2})-\lambda_{i1}k_{xi}(r_{i},\hat{\sigma}_{i})\right]\|\bm{\eta}_{i}\|^{2}\textrm{.}
\end{equation}

Although the dynamic scaling factors $r_{i}$ dominate the nonlinear velocity terms $\frac{k_{r}}{4}c^{2}_{i}|\tilde{\sigma}_{i}|\|\bm{\eta}_{i}\|^{2}$, they are potentially unbounded. Their boundedness can be analyzed by considering the derivative of $V_{r}=\frac{1}{2}\sum_{i=m,s}(r_{i}-c_{ri})^{2}$:
\begin{equation}\label{equ11}
\begin{aligned}
\dot{V}_{r}=&\sum_{i=m,s}(r_{i}-c_{ri})\dot{r}_{i}\\
=&\sum_{i=m,s}\frac{k_{r}c^{2}_{i}r_{i}}{4\lambda_{i1}}(r_{i}-c_{ri})|\tilde{\sigma}_{i}|-\frac{k_{r}}{2}(r_{i}-c_{ri})^{2}\\
\leq&\sum_{i=m,s}\frac{k_{r}c^{4}_{i}r^{2}_{i}}{16\lambda^{2}_{i1}}\tilde{\sigma}^{2}_{i}-\frac{k_{r}}{4}(r_{i}-c_{ri})^{2}\textrm{.}
\end{aligned}
\end{equation}
The projection-based adaptive laws in $\dot{\hat{\sigma}}_{i}$ help dominate $\frac{k_{r}c^{4}_{i}r^{2}_{i}}{16\lambda^{2}_{i1}}\tilde{\sigma}^{2}_{i}$ in \eq{equ11}. From $\sigma_{i}=\|\hbm{x}_{i}\|^{2}$, it follows that 
\begin{align*}
\dot{\sigma}_{i}=2\thbm{x}_{i}\dothatbm{x}_{i}=2\thbm{x}_{i}\fbm{f}_{i}+2k_{xi}(r_{i},\hat{\sigma}_{i})\thbm{x}_{i}\tilbm{x}_{i}\textrm{,}
\end{align*}
and that 
\begin{align*}
\dot{\tilde{\sigma}}_{i}=\theta_{i}-\text{Proj}_{\hat{\sigma}_{i}}(\theta_{i})+2k_{xi}(r_{i},\hat{\sigma}_{i})\thbm{x}_{i}\tilbm{x}_{i}-2k_{\sigma i}(\hbm{x}_{i},r_{i},\hat{\sigma}_{i})\tilde{\sigma}_{i}\textrm{,}
\end{align*}
where $\theta_{i}=2\thbm{x}_{i}\fbm{f}_{i}+2k_{\sigma i}(\hbm{x}_{i},r_{i},\hat{\sigma}_{i})\tilde{\sigma}_{i}$. From~\cite{Stamnes2011Automatica}, the projection operator in \eq{equ3} guarantees that 
\begin{align*}
\Big[\theta_{i}-\text{Proj}_{\hat{\sigma}}(\theta_{i})\Big]\tilde{\sigma}\leq 0\textrm{,}\quad \forall\sigma\geq 0\textrm{,}\ \hat{\sigma}\geq-\epsilon\textrm{.}
\end{align*}
Therefore, the derivative of $V_{\sigma}=\frac{1}{4}\sum_{i=m,s}\tilde{\sigma}^{2}_{i}$ becomes
\begin{equation}\label{equ12}
\begin{aligned}
\dot{V}_{\sigma}=&\frac{1}{2}\sum_{i=m,s}\tilde{\sigma}_{i}\left[\theta_{i}-\text{Proj}_{\hat{\sigma}_{i}}(\theta_{i})\right]\\
&+\sum_{i=m,s}\tilde{\sigma}_{i}\left[k_{xi}(r_{i},\hat{\sigma}_{i})\thbm{x}_{i}\tilbm{x}_{i}-k_{\sigma i}(\hbm{x}_{i},r_{i},\hat{\sigma}_{i})\tilde{\sigma}_{i}\right]\\
\leq&\sum_{i=m,s}\tilde{\sigma}_{i}\thbm{x}_{i}k_{xi}(r_{i},\hat{\sigma}_{i})\tilbm{x}_{i}-k_{\sigma i}(\hbm{x}_{i},r_{i},\hat{\sigma}_{i})\tilde{\sigma}^{2}_{i}\\
\leq&\sum_{i=m,s}\Big\{\frac{1}{k_{r}}\|\bm{\eta}_{i}\|^{2}-k_{\sigma}(\hbm{x}_{i},r_{i},\hat{\sigma}_{i})\tilde{\sigma}^{2}_{i}\\
&\quad \quad \quad +\frac{k_{r}}{4}k^{2}_{xi}(r_{i},\hat{\sigma}_{i})\|\hbm{x}_{i}\|^{2}r^{2}_{i}\tilde{\sigma}^{2}_{i}\Big\}\textrm{.}
\end{aligned}
\end{equation}

\begin{rem}
\normalfont Equations~\eqref{equ11} and \eqref{equ12} show that the observer states $r_{i}$ and $\hat{\sigma}_{i}$ are used to dominate the dynamic nonlinearities due to Coriolis and centrifugal effects. However, because the velocity estimation errors generate potentially destabilizing energy that cannot be dissipated by the P+d controllers through damping injected based on velocity estimates, the designed I\&I observers need to dissipate this energy themselves.
\end{rem}

The summation of $\frac{\alpha_{i}k_{i}}{4}\tildottbm{q}_{i}\tildotbm{q}_{i}$ and 
\eqsthree{equ10}{equ11}{equ12} leads to:
\begin{equation}\label{equ13}
\begin{aligned}
&\dot{V}_{\eta}+\dot{V}_{r}+\dot{V}_{\sigma}+\sum_{i=m,s}\frac{\alpha_{i}k_{i}}{4}\tildottbm{q}_{i}\tildotbm{q}_{i}\\
\leq&-\sum_{i=m,s}\Big\{\psi_{\eta i}\|\bm{\eta}_{i}\|^{2}+\psi_{\sigma i}\tilde{\sigma}^{2}_{i}+\frac{k_{r}}{4}(r_{i}-c_{ri})^{2}\Big\}\textrm{,}
\end{aligned}
\end{equation}
with
\begin{align*}
\psi_{\eta i}=&\lambda_{i1}k_{xi}(r_{i},\hat{\sigma}_{i})-\frac{2}{k_{r}}-\frac{k_{r}}{4}(2\lambda_{i2}+c^{2}_{i}\hat{\sigma}_{i})-\frac{k_{i}}{4}\alpha_{i}r^{2}_{i}\textrm{,}\\
\psi_{\sigma i}=&k_{\sigma i}(\hbm{x}_{i},r_{i},\hat{\sigma}_{i})-\frac{k_{r}}{16}\left(\frac{c^{4}_{i}r^{2}_{i}}{\lambda^{2}_{i1}}+4k^{2}_{xi}(r_{i},\hat{\sigma}_{i})\|\hbm{x}_{i}\|^{2}r^{2}_{i}\right)\textrm{.}
\end{align*}
Note that $V_{o}=V_{\eta}+V_{r}+V_{\sigma}$. For the observer dynamics in \eq{equ3} with suitably selected parameters as in \sect{sec:observer-based controller design}, \eq{equ13} implies that $\dot{V}_{o}\leq-\frac{k_{r}}{2}V_{o}$, and further, that $V_{o}(t)\leq e^{-\frac{k_{r}}{2}t}V_{o}(0)$ and $V_{o}$ globally exponentially converges to zero. Because $r_{i}$ are bounded and $\tilbm{x}_{i}=r_{i}\bm{\eta}_{i}$, it follows that the velocity estimation errors $\tilbm{x}_{i}$ globally exponentially converge to zero themselves.

\begin{rem}
\normalfont \eq{equ13} indicates that the energy generated by the velocity estimation errors is dissipated by the augmented dynamics of the velocity observers, namely the added term $\frac{\alpha_{i}k_{i}r^{2}_{i}}{4\lambda_{i1}}$ in $k_{xi}(r_{i},\hat{\sigma}_{i})$. More specifically, the dynamics of $\hbm{x}_{i}$ in \eq{equ3} behave like filters, i.e., $\dothatbm{x}_{i}=\fbm{f}_{i}+k_{xi}(r_{i},\hat{\sigma}_{i})\tilbm{x}_{i}$. The dynamics of the estimation errors $\tilbm{x}_{i}$ suggest that the augmentations $\frac{\alpha_{i}k_{i}r^{2}_{i}}{4\lambda_{i1}}$ in $k_{xi}(r_{i},\hat{\sigma}_{i})$ increase their speed of convergence. Letting the speed of estimation convergence depend on the dynamic scaling factors limits the energy generated by the estimation errors in a range that the observers can consume.  
\end{rem}

After choosing the P+d control gains to obey
\begin{equation}\label{equ14}
k_{i}\geq\overline{d}_{i}\omega_{i}+\frac{\overline{d}_{j}p^{2}}{4\omega_{j}}+\frac{k_{i}}{\alpha_{i}}\textrm{,}
\end{equation}
where $i,j=m,s$ and $i\neq j$, and combining \eqsa{equ9}{equ13}, the time derivative of $V$ is upper-bounded by $\dot{V}=\sum^{3}_{k=1}\dot{V}_{k}+\dot{V}_{o}\leq 0$. Similar to~\cite{Nuno2008TRO}, $\dot{V}\leq 0$ leads to the conclusion that the teleoperation system is stable: the velocities of, and position error between, the master and slave robots are bounded, i.e. $\{\dotbm{q}_{m},\dotbm{q}_{s},\fbm{q}_{m}-\fbm{q}_{s}\}\in \mathcal{L}_{\infty}$; and the master and slave velocities are square-integrable, i.e. $\{\dotbm{q}_{m},\dotbm{q}_{s}\}\in \mathcal{L}_{2}$, if the inequalities are strict. If, in addition, the hand and environment forces vanish, then $V(t)$ globally asymptotically converges to zero, i.e., the velocities of, and position error between, the master and slave robots asymptotically converge to zero.

\section{Simulations}\label{sec:simulations}

This section illustrates the effectiveness of P+d teleoperation output feedback control based on the augmented I\&I observers through simulations. For simplicity and  without loss of generality, the simulated master and slave robots are identical planar 2-DOF manipulators with revolute joints. The masses and lengths of their links are $m_{1}=1$~kg, $m_{2}=1.5$~kg, $l_{1}=2$~m and $l_{2}=1$~m, and, thus, $\lambda_{i1}=0.3$ and $\lambda_{i2}=20$. The user and environment apply forces along the $y$-axis. The asymmetric time-varying delays $d_{m}$ and $d_{s}$ are upper-bounded by $\overline{d}_{m}=0.2$~s and $\overline{d}_{s}=0.1$~s, respectively. The robots start from $(\fbm{q}_{m}\quad \dot{\fbm{q}}_{m})^\mathsf{T}=(\fbm{q}_{s}\quad \dot{\fbm{q}}_{s})^\mathsf{T}=(\fbm{0}\quad \fbm{0})^\mathsf{T}$ and move under the sinusoidal user-applied force $F_{hy}=4\text{sin}(\pi/20t)+1$~N. The environment with stiffness $k_{e}=1$~kN/m and damping $d_{e}=100$~Ns/m is located at $y_{e}=2$~m.  

After choosing $\omega_{i}=50$, $p=100$ and $\alpha_{i}=4$, the damping gains are selected $k_{i}=20$ to satisfy \eq{equ14}. The observers have parameters $c_{r}=1$, $c_{m}=c_{s}=5$, $k_{r}=5$, $\epsilon=0.1$, and initial states $r_{0}=2$, $\hbm{x}_{0}=(0.05\quad 0.02)^\mathsf{T}$, $\hat{\sigma}_{0}=\|\hbm{x}_{0}\|^{2}$ and $\bm{\xi}_{0}=\hbm{x}_{0}$. 

\begin{figure}[!hbt]
\centering
\includegraphics[width=8cm,height=5cm]{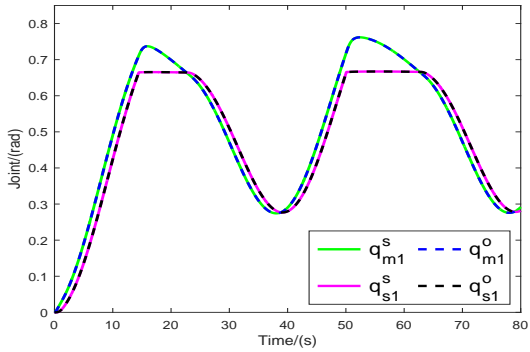}
\caption{The position of the first robot joints for state feedback P+d control~(master $q^{s}_{m1}$ and slave $q^{s}_{s1}$), and for the proposed output feedback control~(master $q^{o}_{m1}$ and slave $q^{o}_{s1}$).}
\label{fig1}
\end{figure}

\begin{figure}[!hbt]
\centering
\includegraphics[width=8cm,height=5cm]{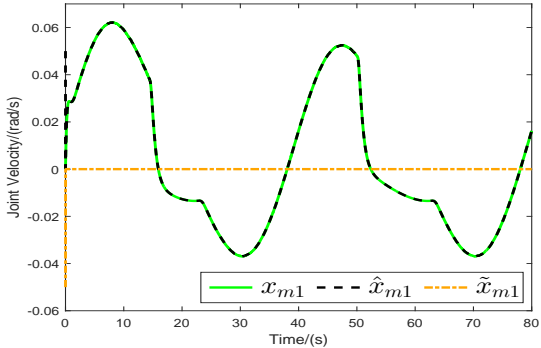}
\caption{The actual velocity $x_{m1}$, estimated velocity $\hat{x}_{m1}$ , and estimation error $\tilde{x}_{m1}$ for the first joint of the master robot.}
\label{fig2}
\end{figure}

\begin{figure}[!hbt]
\centering
\includegraphics[width=8cm,height=5cm]{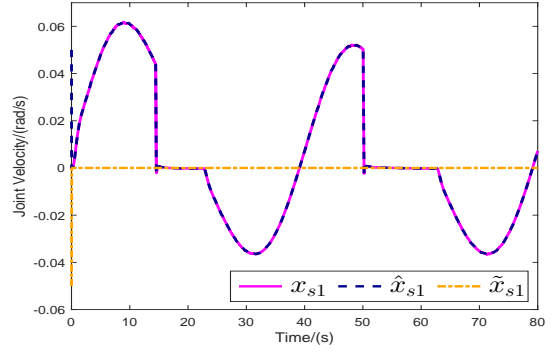}
\caption{The actual velocity $x_{s1}$, estimated velocity $\hat{x}_{s1}$ , and estimation error $\tilde{x}_{s1}$ for the first joint of the slave robot.}
\label{fig3}
\end{figure}

To save space, \figs{fig1}{fig3} show the numerical results only for the first joints of the two simulated robots. In \fig{fig1}, the positions of first joint of the master and slave robots under the conventional state feedback P+d control and under the proposed output feedback P+d control almost overlap. The slave tracks the position of the master in free motion. When the master moves into the environment, $\{q^{s}_{m1},\ q^{o}_{m1}\}>0.65$~rad, the slave stops in contact with the environment and the master continues to move forward as allowed by the proportional gain of the P+d controller. In \figs{fig2}{fig3}, the master and slave velocity estimates~($\hat{x}_{m1}$ and $\hat{x}_{s1}$) converge to the actual velocities~($x_{m1}$ and $x_{s1}$) right away; and the estimation errors remain close to zero even for relatively large initial velocity estimation errors~($0.05$~rad/s). The simulation results illustrate that the proposed output feedback synchronization strategy has similar performance to state feedback P+d control. 

Compared to the I\&I observer in~\cite{Stamnes2011Automatica}, the main advantage of the new constructive $\text{I\&I}$ observers is the simpler analytical solution of their designed dynamics. First, the new observers have only $n+2$ states, instead of $2n+2$ states, because they eliminate the need for position estimates $\hat{\fbm{y}}_{i}$. Second, they require no analytical computation of state transformations because they only need the inverses inertia matrices $\ibm{M}_{i}(\fbm{y}_{i})$ which can be computed numerically. Third, they simplify the $\dotfbm{\xi}$ dynamics by replacing the $n\times n$ matrix gain $\fbm{K}_{x}(\hat{\sigma},\hbm{y})$ with a scalar gain $\dot{k}_{xi}(r_{i},\dot{r}_{i},\dot{\hat{\sigma}})$, and by replacing the non-smooth functions $\bar{\Delta}_{y}(\fbm{y},\hbm{y},\hat{\sigma})\|\tilbm{y}\|+\bar{\Delta}_{\sigma}(\fbm{y},\hbm{x},\hat{\sigma})\|\tilde{\sigma}\|$ with $\frac{k_{r}}{4\lambda_{i1}}c^{2}_{i}|\tilde{\sigma}|$.

\section{Conclusions}\label{sec:conclusions}

This paper has proposed a globally stable output feedback control strategy for nonlinear bilateral teleoperation systems with time-varying delays. The strategy uses simplified and augmented constructive I\&I observers at the master and slave sides to dispense with velocity measurements. The new observers do not estimate positions and increase the speed of convergence of the velocity estimates through dynamic scaling factors. Based on Lyapunov stability analysis, the paper has derived design criteria for the observer and controller parameters that guarantee globally stable teleoperation under the proposed output feedback P+d control strategy. Numerical simulations have verified that the new observer-based P+d controller stabilizes nonlinear teleoperation systems with time-varying delays without using velocity measurements, and achieves position tracking performance similar to that of conventional state feedback P+d control. Because the proposed output feedback controller is model-based, upcoming research will investigate globally adaptive output feedback control approaches to make the design robust to system uncertainties.

\bibliography{bibi}
\end{document}